\documentclass[10pt,sort&compress,3p,onecolumn]{elsarticle}

\usepackage[normalem]{ulem}
\usepackage{array} 
\usepackage{paralist} 
\usepackage{verbatim} 
\usepackage{helvet}
\usepackage{amssymb}
\usepackage{mathrsfs}
\usepackage{graphicx}
\usepackage{subfig}
\usepackage[sumlimits]{amsmath}
\usepackage{braket} 
\usepackage{nicefrac} 
\usepackage{accents} 
\usepackage{bbm} 
\usepackage{textcomp} 
\usepackage{booktabs} 

\journal{Journal of Magnetic Resonance}

\begin{document}

\begin{frontmatter}

\title{Bandwidth-Limited Control and Ringdown Suppression in High-Q Resonators}
\author[mit,iqc]{Troy W. Borneman\corref{cor1}\fnref{fn1}}
\ead{troyb@mit.edu}

\author[mit,iqc,pi,uwchem]{David G. Cory}

\cortext[cor1]{Corresponding author.}
\fntext[fn1]{Present address: Institute for Quantum Computing, 200 University Ave. W., Waterloo, ON, Canada N2L 3G1}

\address[mit]{Department of Nuclear Science and Engineering, Massachusetts Institute of Technology\\
                  Cambridge, MA, USA}
\address[iqc]{Institute for Quantum Computing, Waterloo, ON, Canada}
\address[pi]{Perimeter Institute for Theoretical Physics, Waterloo, ON, Canada}
\address[uwchem]{Department of Chemistry, University of Waterloo, Waterloo, ON, Canada}


\begin{abstract}
We describe how the transient behavior of a tuned and matched resonator circuit and a ringdown suppression pulse may be integrated into an optimal control theory (OCT) pulse-design algorithm to derive control sequences with limited ringdown that perform a desired quantum operation in the presence of resonator distortions of the ideal waveform. Inclusion of ringdown suppression in numerical pulse optimizations significantly reduces spectrometer deadtime when using high quality factor (high-Q) resonators, leading to increased signal-to-noise ratio (SNR) and sensitivity of inductive measurements. To demonstrate the method, we experimentally measure the free-induction decay of an inhomogeneously broadened solid-state free radical spin system at high Q. The measurement is enabled by using a numerically optimized bandwidth-limited OCT pulse, including ringdown suppression, robust to variations in static and microwave field strengths. We also discuss the applications of pulse design in high-Q resonators to universal control of anisotropic-hyperfine coupled electron-nuclear spin systems via electron-only modulation even when the bandwidth of the resonator is significantly smaller than the hyperfine coupling strength. These results demonstrate how limitations imposed by linear response theory may be vastly exceeded when using a sufficiently accurate system model to optimize pulses of high complexity.  
\end{abstract}


\end{frontmatter}

\section{Introduction}
Inductive coupling to a tuned and matched resonator circuit is a standard technique for the manipulation and measurement of a quantum system. In a common application, a resonator converts a pulsed electrical signal into a pulsed magnetic field and the resulting magnetic field produced by the quantum system back into an electrical signal. The same principles apply as well to capacitive coupling to electric fields. The conversion efficiency is determined by a number of factors, including the quality factor (Q) of the resonator, defined as the ratio of stored to dissipated energy in the resonant circuit. All other factors being equal, both the magnetic field strength resulting from the application of an electrical signal of a certain power and the signal-to-noise ratio (SNR) of an electrical signal induced by the magnetic response of the sample scale as $\sqrt{Q}$ \cite{SchweigerBook,AbragamBook,Freed1989,Rinard:2004a}.

An important consequence of using tuned circuits for pulse transmission is that any resonator has an associated ringdown time during which energy stored in the circuit dissipates. The signal to be observed from the quantum system is typically many orders of magnitude smaller than the control amplitudes, normally requiring a spectrometer deadtime of at least five times the ringdown time before receiver circuitry, such as low-power high-gain preamplifiers, can be switched on. When moving to a high-Q resonator, to enable faster control and greater SNR, the relative amount of stored energy in the circuit increases, leading to longer ringdown and a deadtime that can exceed the characteristic phase coherence time ($T_2$) of the quantum signal, significantly lowering measurement SNR and possibly preventing observation of the quantum signal at all. A dispersion of interaction strengths between quantum degrees of freedom in the sample may also lead to a spectral breadth which exceeds the resonator bandwidth of frequencies which are efficiently converted, preventing the rotation or detection of the entire spectrum with a single unoptimized pulse \cite{Hornak1986}. 

Another challenge inherent to the use of resonant circuits for pulse transmission is the distortions they produce of an ideal pulse waveform, once again due to the finite response time of the reactive elements used to construct a tuned circuit. For low values of Q (roughly 10 - 200), the effects of waveform distortions are relatively minor, resulting in a typical drop in pulse fidelity of a few percent \cite{HodgesThesis}. However, as the Q is increased, the pulse action experienced by the quantum system differs drastically from the ideal response of an undistorted waveform. For these reasons it is common practice, especially in the field of pulsed electron spin resonance (ESR), to intentionally spoil the Q to reduce it to a value where these effects are no longer significant \cite{Rinard1994}. 

There are many applications of pulsed induction techniques that benefit from the high sensitivity provided by high-Q resonators. For example, thin film samples have a number of well-known applications in chemistry and materials science and have drawn recent interest in the field of quantum information processing (QIP), where a highly structured 2D geometry maps naturally to node-based proposals for quantum computing \cite{Lloyd:1993a,Mehring:2006a,Cappellaro:2009a,Yao:2012a,Borneman:2012a}. These samples are particularly well-suited for study with high-Q, large filling factor resonators \cite{Malissa:2012a,Benningshof:2012a}. Additionally, advances in inductive imaging techniques have allowed for sub-micron resolution, pushing the limits of sensitivity using conventional resonators \cite{Twig:2011a,Shtirberg:2011a,Suhovoy:2009a}. As more progress is made on the development of novel quantum devices the need for high sensitivity measurement will increase, requiring further development of control techniques with resonators of increasingly high Q.

Integration of the resonator impulse response into an optimal control theory (OCT) pulse-design algorithm has been suggested as a means of accounting for resonator distortions to obtain increased pulse fidelity in inductively controlled quantum systems \cite{Motzoi:2011a}. This technique was recently applied in pulsed ESR to accurately tailor spin response when using a high-Q resonator \cite{Spindler:2012a}. We extend these results by discussing how a ringdown suppression pulse may also be included in OCT pulse-design algorithms. We demonstrate our suggested method by optimizing and implementing a bandwidth-limited OCT pulse, robust to static and microwave field inhomogeneity, that allows the direct observation of the free-induction decay (FID) of an inhomogeneously broadened sample of irradiated fused-quartz \cite{Eaton1993,Eaton2010} in a high-Q ($\approx$ 10,000) X-band ($\approx$ 10 GHz) rectangular cavity. We also discuss the application of high-Q pulse optimization techniques to achieving universal control of a solid-state electron-nuclear spin system via electron-only modulation even when the resonator bandwidth is significantly less than the hyperfine coupling strength.

We begin by reviewing a general model of the transient response of a resonator and methods of ringdown suppression by application of a phase-inverted compensation pulse. We then show how the resonator impulse response function and ringdown suppression may be integrated into an OCT algorithm for the design of universal rotation pulses. We end with a discussion of how limitations imposed by linear response theory may be vastly exceeded when using complex pulses numerically optimized using a sufficiently accurate system model, with particular emphasis on solid-state electron-nuclear spin systems.

\section{Resonator Model and Ringdown Suppression}
\label{Sec:ResonatorModel}
\begin{figure*}[p]
 \centering
 \includegraphics[width=12cm]{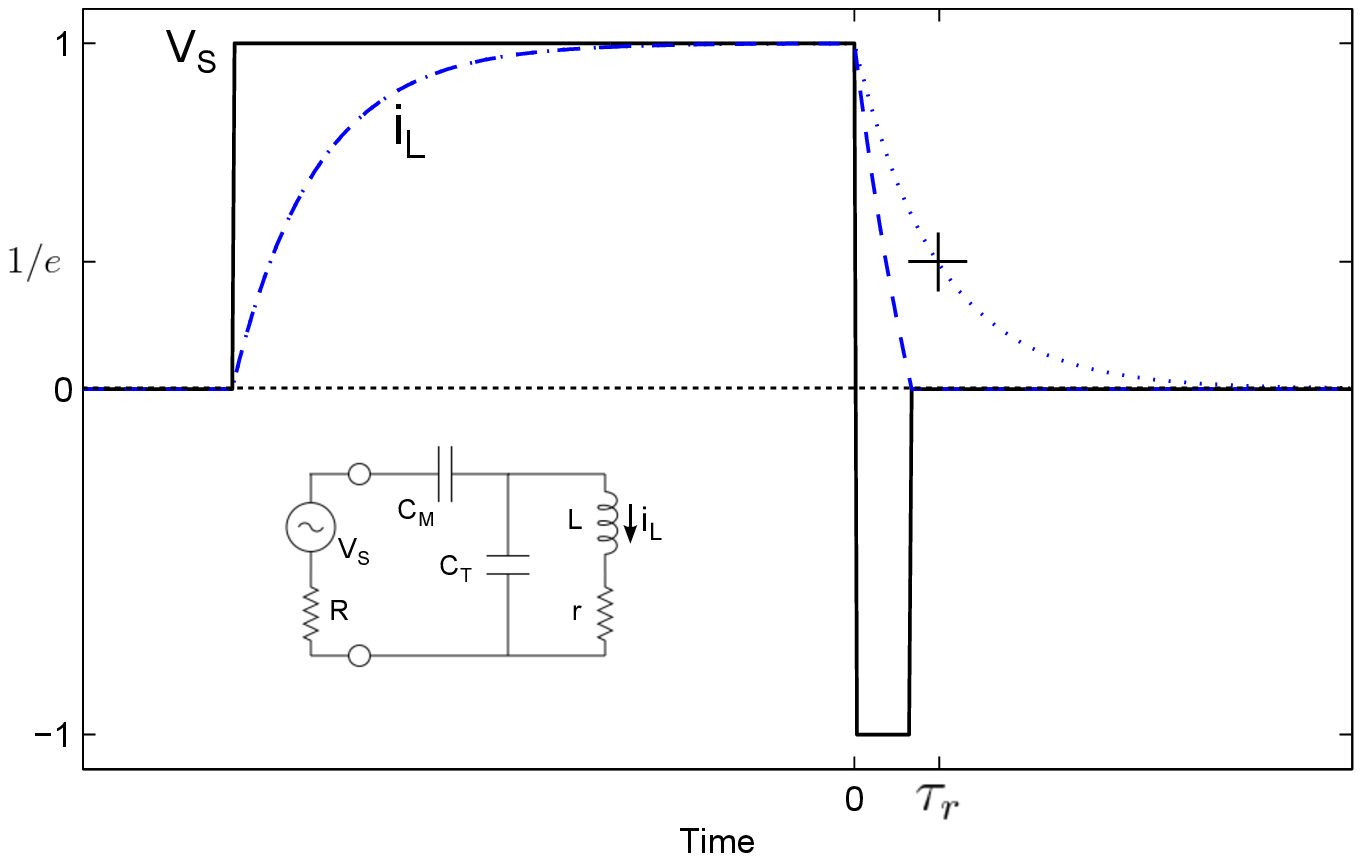}
 \caption{\textbf{Transient behavior of a resonator circuit.} A general resonant transmission circuit may be modeled as a series tuned RLC circuit capacitively coupled to a time-varying voltage source \cite{Barbara:1991a} (inset). The quality factor (Q) of the resonator is given mainly by the circuit resonance frequency, $\omega_0 = 1/\sqrt{L C_T}$, the coil inductance,$L$, and the coil resistance, $r$: $Q\approx \omega_0 L/r$. The circuit impedance is matched to $R_0 = 50$ Ohms by varying the capacitances $C_T$ and $C_M$. For high-Q resonators, the dominant transient response of the resonator to a square pulse input of voltage, $V_S$, (bold line) is an exponential rise (dashed line) and subsequent ring-down (dotted line) of the coil current, $i_L$, and resulting magnetic field. The ringdown may be suppressed by application of a phase inverted compensation pulse at the end of the square pulse to drive the coil current to zero. The characteristic transient ringdown time without compensation, $\tau_r = Q/\omega_0$, is denoted by a cross.}
 \label{fig:ResonatorCircuit}
\end{figure*}

Several models of the transient behavior of a resonator have been suggested and analyzed in detail \cite{Mehring1972,Barbara:1991a,Takeda2009}, with various methods suggested for the suppression of pulse transients \cite{Vaughan:1972a,Zhang:1990a,Vega:2004a,Tabuchi2010}. We use the model of Barbara, \textit{et al.} \cite{Barbara:1991a} consisting of a series RLC resonant circuit capacitively coupled to a voltage source through a real 50 Ohm impedance (Figure \ref{fig:ResonatorCircuit}). We restate here the results of their analysis relevant to our integration of the model into an OCT algorithm, with further details available in the references.     

The reactive elements necessary for efficient signal conversion have a finite response time to variations of the control signal, causing distortions of the ideal waveform. For a linear, time-invariant system, the waveform incident upon the quantum system, $y(t)$, is given by the convolution of the ideal control pulse, $x(t)$, with the impulse response, $h(t)$, of the transmitting circuit,
\begin{equation}
y(t) = h(t) \ast x(t) = \int_{-\infty}^{\infty}h(t-\tau)x(\tau) \, d\tau.
\label{eq:ConvolutionIntegral}
\end{equation}
The impulse response gives a complete description of the transient behavior of the transmitting circuit and may be either inferred from measurement, or derived from an appropriate model. The ideal pulse is described as a time varying voltage of amplitude $v(t)$ and phase $\phi(t)$ applied at a driving frequency, $\omega_t$:
\begin{equation}
V_s(t) = v(t)\cos{(\omega_t t+\phi(t))}.
\end{equation}
When the driving frequency is set near resonance with energy level splittings of the quantum system, the magnetic field induces transitions between the eigenstates of the system Hamiltonian. This model is representative of a general quantum control scenario and allows us to analytically understand the dynamics of high-Q resonators. 

The resonator circuit response to an impulse of constant amplitude and phase may be determined by the application of Laplace transformation techniques \cite{LaplaceTransformBook}. The resulting filtered output waveform in the time domain is determined by the inverse Laplace transform, calculated via a partial fraction expansion, of the filtered waveform in the $s$-domain. The transient coil current, $i_L(t)$, which is proportional to the magnetic field applied to the sample, may be represented as a sum of poles, $\delta_k$, and corresponding residues, $d_k$, after inverse Laplace transformation:
\begin{equation}
i_L(t) \propto \sum_k d_k e^{-\delta_k t}.
\end{equation}

The four reactive elements in the circuit lead to four poles of the transfer function:
\begin{enumerate}
\item A steady-state oscillation at the driving frequency, $\omega_t$.
\item An exponentially decaying DC transient.
\item An exponentially decaying oscillation at the probe free-ringing frequency, $\Omega$.
\item An exponentially decaying oscillation at minus the probe free-ringing frequency, -$\Omega$.
\end{enumerate}
It is the third and fourth poles that are of most interest for determining the transient response of the circuit, as the DC transient is normally very small and the steady-state oscillation may be eliminated by moving into a reference frame rotating at the driving frequency. In a rotating wave approximation, this frame rotation also allows us to neglect the effect of the fourth pole. 

The relevant third pole, $\delta_3$, may be written as
\begin{equation}
\delta_3 = \delta_4^{\ast} = \gamma - i\omega_0 f\left(1-\sqrt{\frac{r}{4R_0 f^2}}\right),
\end{equation}      
where $\gamma = \omega_0/Q$ is the rate of transient decay in terms of the inductor quality factor, $Q=\omega_0 L/r$, and 
\begin{equation}
f = \sqrt{1-\frac{1}{4Q^2}}
\end{equation}
is a scaling factor determining how close the resonator free-ringing frequency, $\Omega$, is to the tuned frequency, $\omega_0 = \sqrt{1/LC_T}$. For high-Q resonators, $r<<R_0$, $\Omega \approx \omega_0$, and the transient oscillations in the rotating frame are critically damped. The resulting dynamics may then be approximated as a simple exponential ringup and ringdown of the pulse amplitude with time-constant
\begin{equation}
\tau_r = Q/\omega_0.
\end{equation}
This time-constant differs from the commonly used $2Q/\omega_0$, which is valid only for an isolated series RLC circuit without the inclusion of a matching capacitor \cite{Mehring1972}. Such approximations are used here only for the sake of argument and demonstration; when high-fidelity control of a particular system is desired the full form of the resonator transfer function must be experimentally determined \cite{Spindler:2012a}.

Resonator ringdown may be suppressed by the application of a trailing compensation pulse of appropriate length and amplitude to drive the energy stored in the resonator to zero at the end of the pulse \cite{Hoult1979,MehringBook}. As shown in Figure \ref{fig:ResonatorCircuit}, this compensation pulse can significantly shorten the spectrometer deadtime, but at the expense of introducing an additional rotation to the quantum system. In the next section, we describe how the resonator transfer function and a ringdown compensation pulse may be integrated into an optimal control theory (OCT) algorithm to enable high-fidelity operations in the presence of transient effects and the additional rotation introduced by ringdown suppression.

\section{Optimal Control Theory for High-Q Resonators}

In the absence of dissipative processes, the problem of control sequence design may be stated as optimizing a time-dependent Hamiltonian, $H(t)$, such that the action of the Hamiltonian on a general quantum state, $\rho$, yields the desired dynamics, as governed by the Liouville-von Neumann equation,
\begin{equation}
\frac{d\rho}{dt}=-\frac{i}{\hbar}\left[\rho,H(t)\right].
\end{equation}
As is common practice for dynamical calculations, we will use the convention $\hbar=1$ for the remainder of this work. The formal solution of this equation may be written as
\begin{equation}
\rho(t) = U(t)\rho(0)U^{\dag}(t),
\end{equation}
where $U(t)$ is a unitary propagator representing time evolution under the applied Hamiltonian:
\begin{equation}
U(t) = \mathcal{T}e^{-i\int_0^t{H(s)\,ds}},
\label{eq:UnitaryProp}
\end{equation}
where $\mathcal{T}$ is the Dyson time-ordering operator accounting for non-commumitivity of $H(t)$ with time. For this work, we are interested in finding a time-dependent Hamiltonian that accurately implements a desired unitary operation, $U_{\mathrm{d}}$. The notion of accuracy for unitary operations is quantified by defining an appropriate performance functional, such as the average gate fidelity for Hilbert space dimension $D$:
\begin{equation}
\Phi(U_{\mathrm{d}},U_{\mathrm{pulse}}) = \left|\left<U_{\mathrm{d}}|U_{\mathrm{pulse}}\right>\right|^2 =
\frac{\left|\mathrm{Tr}\left[U_{\mathrm{d}}^{\dag}U_{\mathrm{pulse}}\right]\right|^2}{2^D}.
\label{eq:AverageGateFidelity}
\end{equation} 
This performance functional has been shown to be equivalent to the average state fidelity over a complete set of input states \cite{Fortunato2002} and guarantees the desired evolution over all possible input states when $\Phi$ is close to unity.

The set of unitary operations that may be generated depends on the form of the accessible physical interactions present in a given system. These interactions are commonly separated into drift Hamiltonian terms, $H_d$, which are time-independent, and control Hamiltonian terms, $H_c(t)$, which may be varied during the course of an experiment. We parameterize the control Hamiltonians into a set of constant Hamiltonians, $\{H_k\}$, for which we may independently vary amplitudes, $u_k(t)$, to yield a total system Hamiltonian of
\begin{equation}
H(t) = H_d + \sum_k u_k(t) H_k.
\end{equation} 

To avoid having to evaluate a set of time-ordered exponentials to solve (\ref{eq:UnitaryProp}), it is common practice to discretize the time-dependence of the control amplitudes into $N$ intervals of length $t_j$, such that the unitary propagator for a time $T = \sum_{j=1}^N{t_j}$ may be formally expressed as a product of time-independent exponentials:
\begin{equation}
U(T) = \prod{U_j} = \prod_{j=1}^N{e^{-i(H_d+\sum_k{u_k^j H_k})t_j}}.
\end{equation}
For the sake of simplicity, we will assume $t_j = \Delta t$ for all $j$. The validity of this time-slicing method of propagator evaluation depends on the chosen value of $\Delta t$, with accuracy increasing for smaller time discretization intervals. With all time-dependence effectively removed from the problem, the control objective now simplifies to determining the set of constant amplitudes, $\{u_k^j\}_{j=1}^N$, which maximize the value of the average gate fidelity (\ref{eq:AverageGateFidelity}), which may be achieved through application of optimal control theory (OCT) \cite{PontryaginBook}. In the following, we utilize the efficient GRadient Ascent Pulse Engineering
(GRAPE) OCT algorithm \cite{GRAPE2005}, which has been used extensively in spin-based implementations of quantum information processing (see \cite{Henry2007,Ryan2008,Moussa2011} for example applications of the GRAPE algorithm). This method is limited to finding solutions that correspond to local optima, but high-fidelity pulses may still be consistently found, insensitive to the initial guess, that are useful in practice.

\subsection{Optimizing Infinite-Bandwidth Controls}

The relative efficiency of the GRAPE algorithm is derived from the need to only compute the propagator corresponding to each time step, $U_j$, a single time at each iteration of the optimization. The resulting single iteration propagators are then used to calculate both the value of the performance functional and an approximation of gradients used for pulse updating. The truncation of the gradients is valid when $\Delta t$ is chosen to be significantly smaller than the inverse magnitude of the system Hamiltonian. Approximating the gradients in this way avoids the use of finite-differencing and other inefficient methods of computing the gradients. In the notation of \cite{GRAPE2005}, the average gate fidelity and corresponding gradients to first order in $\Delta t$ are
\begin{equation}
\Phi = \left|\left<U_{\mathrm{d}}|U_{\mathrm{pulse}}(T)\right>\right|^2 = \left<P_j|X_j\right>\left<X_j|P_j\right>
\end{equation}
and
\begin{equation}
\frac{\delta\Phi}{\delta u_k^j} = -2\mathrm{Re}\left[\left<P_j|i\Delta t H_k X_j\right>\left<X_j|P_j\right>\right],
\end{equation}
where $P_j=U^{\dag}_{j+1}...U^{\dag}_N U_d$ and $X_j=U_j...U_1$ are the backward and forward representations, respectively, of the total pulse propagator at the $j^{\mathrm{th}}$ time interval. 

The control amplitudes at optimization iteration $n$, $u_k^j(n)$, are updated according to
\begin{equation}
u_k^j(n) = u_k^j(n-1) + \epsilon\frac{\delta\Phi}{\delta u_k^j(n-1)},
\label{eq:UpdateRule}
\end{equation}
where $\epsilon$ is an adjustable scaling factor for the gradients to prevent hindering algorithm convergence by making large deviations in the control parameters between iterations. The algorithm proceeds as follows:
\begin{enumerate}
\item An initial set of control parameters is defined: $\{u_k^j\}_{j=1}^{N}(0)$
\item The piecewise constant propagators for each time-step are calculated: $U_j$
\item The value of the performance functional is calculated: $\Phi$
\item If $\Phi \geq \Phi_{\mathrm{targ}}$ the algorithm exits, otherwise the gradients for each time-step and control are calculated: $\frac{\delta\Phi}{\delta u_k^j}$
\item A line-search is used to optimize the step size, $\epsilon$, in the gradient direction
\item The control parameters are updated according to (\ref{eq:UpdateRule})
\item Loop to Step 2
\end{enumerate}
To include ringdown suppression and account for distortions of the ideal control parameters due to the limited bandwidth of a high-Q resonator we make several modifications to this base algorithm, similar to those previously suggested \cite{Motzoi:2011a}.

\subsection{Optimizing Bandwidth-Limited Controls}
\label{Sec:DistortedControls}
We first resample the control parameters to ensure the calculated propagators and gradients accurately reflect the distortions of the ideal control fields during periods when the applied control voltages are constant. For clarity, we define a \textit{control period} as the time $\Delta t$ during which the applied control voltage is constant, and a \textit{evolution period} as the resampled time $\Delta t/n_s$ during which the control field seen by the quantum system is approximated as constant. Concretely, we begin with a set of undistorted controls, $\{u_k^j\}$, where $j = 1,...,N$ and resample to obtain a new set of undistorted controls $\{\widetilde{u}_k^m\}$, where $m = 1,...,M$ and $M=n_s N + n_r$ for $n_s$ samples per $N$ control periods and $n_r$ values of zero appended to the waveform to account for pulse ringdown. We then define a mapping between the resampled undistorted controls and a new set of distorted controls, $\{\widetilde{v}_k^m\}_{m=1}^M$, given by the discrete convolution of the undistorted controls with the resonator transfer function (\ref{eq:ConvolutionIntegral}):
\begin{equation}
\widetilde{v}_k^m = h_k \ast \widetilde{u}_k^m = \sum_{l\leq m}{h_k^{m-l+1}\widetilde{u}_k^l}.
\end{equation}

In contrast to the infinite-bandwidth case, the value of the control parameters during the $m^{\mathrm{th}}$ evolution period depends to a certain degree on the value of all previous controls. As a result, the gradient of the performance functional when control period $j$ is perturbed depends on the change in the unitary propagator for all evolution periods $m > (j-1)n_s$. As derived in the Appendix, the gradients for bandwidth-limited controls are
\begin{equation}
\frac{\delta\Phi}{\delta u_k^j} = \sum_{m > (j-1)n_s}^M \xi_k^m(j) \frac{\delta\Phi}{\delta \widetilde{v}_k^m}, 
\end{equation}
where $\xi_k^m(j)$ is the convolution of a top-hat function, $\Xi$, with the resonator response:
\begin{equation}
\xi_k^m(j) = \sum_{l \leq m} h_k^{m-l+1} \Xi(j,n_s).
\end{equation}
The top-hat function, which is formally defined in the Appendix, accounts for the resampling of the control periods into evolution periods. The convolution of this function with the resonator impulse response may be interpreted as a weighting function of the gradients of all evolution periods affected by the perturbation of a given control period. The individual gradients of the evolution periods are
\begin{equation}
\frac{\delta\Phi}{\delta \widetilde{v}_k^m} = -2\mathrm{Re}\left[\left<\widetilde{P}_m|i\frac{\Delta t}{n_s} H_k \widetilde{X}_m\right>\left<\widetilde{X}_m|\widetilde{P}_m\right>\right],
\end{equation} 
where $\widetilde{X}_m$ and $\widetilde{P}_m$ are defined in the Appendix. Note that pulse updating is done only for control periods, with evolution periods serving as a calculational tool. 

\begin{figure*}[p]
 \centering
 \includegraphics[width=16cm]{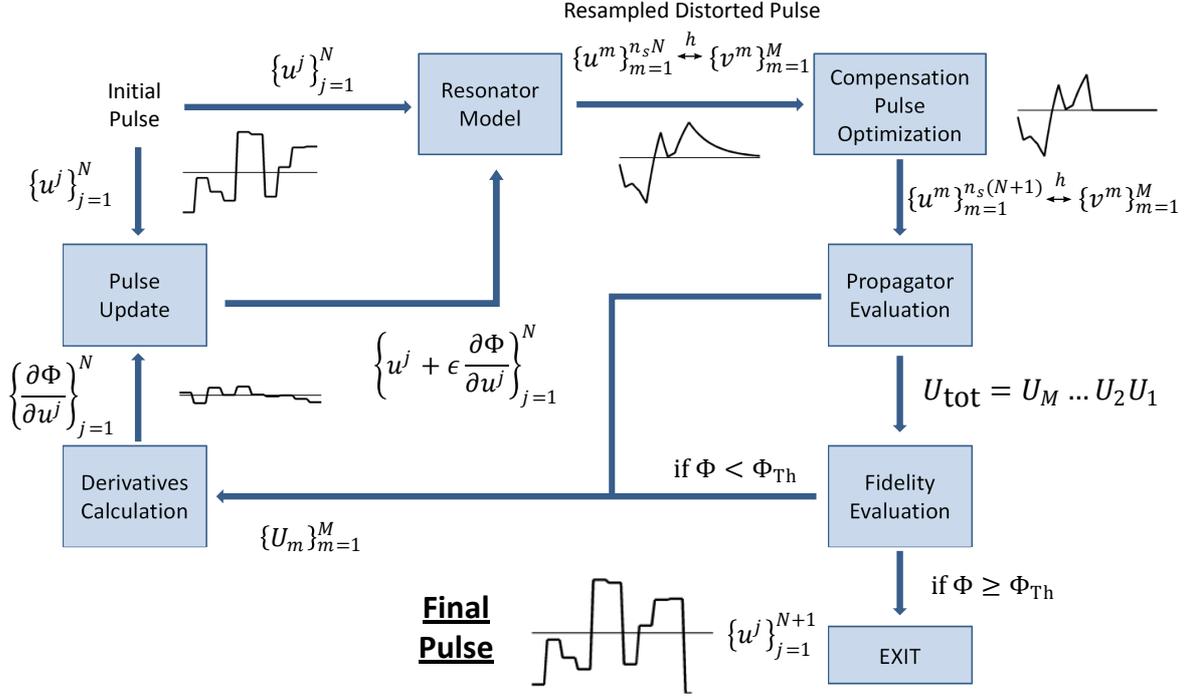}
 \caption{\textbf{Schematic of optimization algorithm.} The bandwidth-limited GRAPE pulse optimization algorithm proceeds in a similar manner as for undistorted controls \cite{GRAPE2005}, with the resonator transfer function included in the calculation of the pulse propagator, average gate fidelity, and gradients. The notation is explained in the main text, with $k=1$ for simplicity. An initial guess of control amplitudes is resampled and convolved with the resonator transfer function to yield a distorted set of control amplitudes including ringdown. A compensation pulse period is then optimized in a sub-routine and appended to the waveform to yield a distorted set of control amplitudes with minimized ringdown. The propagator for each evolution period of the distorted control amplitudes is then calculated and the average gate fidelity computed. The set of propagators is used to calculate the gradient direction of the fidelity with respect to the undistorted controls. A line search is then performed to optimize the step-size in the gradient direction. At each step of the line-search the sub-routines to calculate the updated distorted control amplitudes and corresponding compensation pulse are called, accounting for the compensation pulse not being included in the gradient calculation. The undistorted controls are then updated according to the gradient direction and step-size and the process iterates until a desired value of the fidelity is achieved.}
 \label{fig:OptimizationAlgorithm}
\end{figure*}

Optimization of the compensation pulse for elimination of ringdown is performed as a sub-routine and is not considered in the calculation of the gradient direction. However, the compensation pulse is taken into account while calculating the total pulse fidelity and while performing a line search to optimize the step size in the gradient direction, $\epsilon$. In practice, the line search is performed by choosing three values of $\epsilon$, optimizing the compensation pulse for each value, evaluating the fidelities of the resulting pulses, fitting to a quadratic, and taking the maximum value. The routine for the compensation pulse optimization may be implemented as a two-dimensional search over the length and amplitude of an appended final control period that minimizes the ringdown. An implementation of the GRAPE algorithm for bandwidth-limited control is described in Figure \ref{fig:OptimizationAlgorithm}.

\section{Observation of Free-Induction Decay at High-Q}
We performed experiments using a standard irradiated fused-quartz sample contained in a Varian E-231 rectangular cavity on a home-built X-band pulsed ESR spectrometer. The loaded quality factor and impulse response function of the cavity were measured by fitting an exponential to the rising and falling edges of a square pulse digitized with a pick-up coil inserted into a cavity iris opposite to the sample. Care was taken to only weakly couple the pick-up coil to the cavity fields in order to disturb the cavity mode structure as little as possible. 

We measured a loaded $Q$ of 8,486 for the Varian cavity, giving a ringdown time-constant of $\tau_r = 142$ ns at a resonance frequency of $\omega_0/2\pi = 9.5236$ GHz. The $r^2$ value of the exponential fit was 0.9804, verifying that modeling the transient behavior of our high-Q resonator as an exponential rise and fall of the field strength is a good approximation. We found that a spectrometer deadtime of 1.2 $\mu$s was required to allow the pulse ringdown to decay to a value below the spectrometer noise floor. The inhomogeneous phase coherence relaxation time of the sample, measured by observing a spin echo, was $T_2^* \approx 250$ ns, preventing us from observing a useful FID of the quartz sample when using an unoptimized pulse. 

To observe the quartz FID we optimized a bandwidth-limited OCT pulse that performs a $\pi/2$ rotation about the x-axis robust to variations in the static and microwave field strengths. The resonator model discussed in Section \ref{Sec:ResonatorModel} was included in the optimization, as well as ringdown suppression. The Hamiltonian used for optimization, in a frame rotating at the nominal electron Larmor frequency, was
\begin{equation}
H(\Delta\omega,\omega_1) = \frac{1}{2}\Delta\omega \sigma_z + \frac{1}{2}\omega_1 A(t) \sigma_x,
\label{eq:SingleSpinHam}
\end{equation} 
where $\Delta\omega$ is a resonance offset parameter representing static field inhomogeneity in units of rad/s, $\omega_1$ is a scaling factor of the nominal Rabi frequency, $\omega_{1,\mathrm{nom}}$, representing microwave field inhomogeneity, and $A(t)$ is the time-dependent amplitude modulation representing the OCT pulse, with range $\{-\omega_{1,\mathrm{nom}},\omega_{1,\mathrm{nom}}\}$. The spin dynamics was calculated by taking a convex operator sum over a uniform classical probability distribution of field inhomogeneities, $P(\Delta\omega,\omega_1)$ \cite{Borneman:2010a}.

\begin{figure*}[p]
 \centering
 \subfloat[Time-Domain Profile]{\label{fig:OptimizedPulse}\includegraphics[width=12cm]{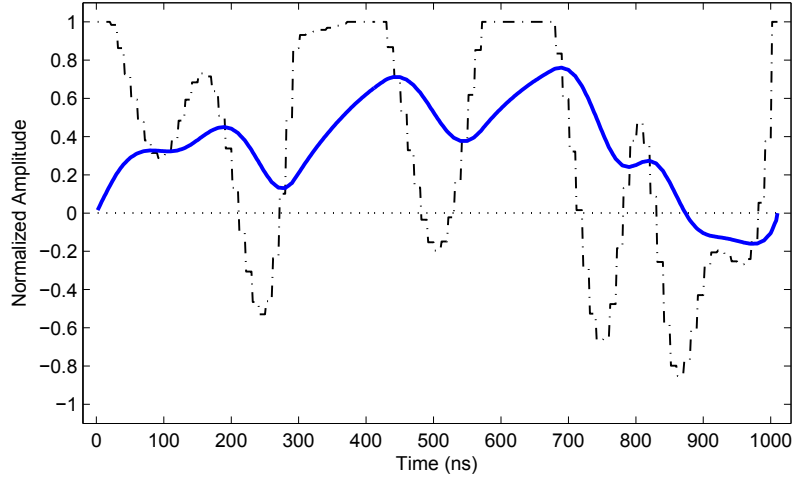}}
 \\
 \subfloat[Simulated Pulse Response]{\label{fig:OptimizedPulseResponse}\includegraphics[width=12cm]{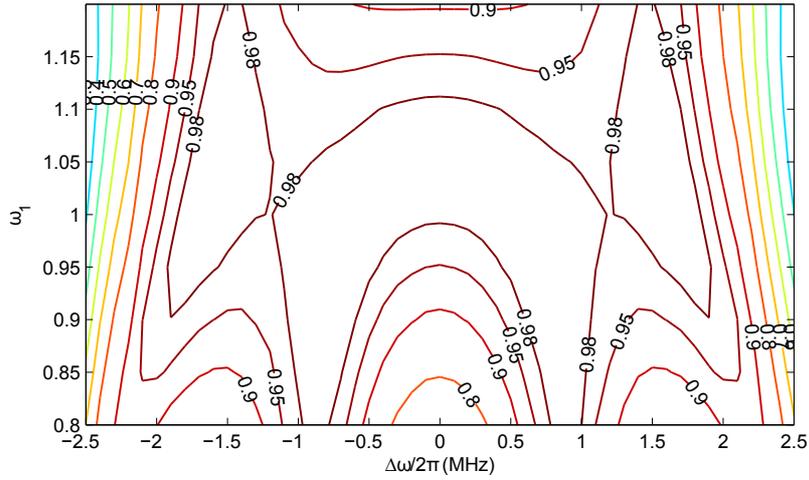}}
 \caption{\textbf{Bandwidth-limited OCT pulse robust to static and microwave field inhomogeneities.} (a) Time-domain profile of a transients optimized OCT pulse with ringdown suppression implementing a $\pi/2)_x$ rotation robust to variations in the static and microwave field strengths. The dashed line represents the undistorted controls and the solid line represents the control fields seen by the spin system after transmission through a resonator with Q = 8,486. The control amplitudes are normalized to a nominal Rabi frequency of $\omega_{1,\mathrm{nom}}/2\pi=5.26$ MHz. Optimization parameters and further details are discussed in the main text. Note that this pulse is phase-refocused, in that all spins in the sample are rotated with the same phase. (b) Simulated pulse fidelity over an extended range of static and microwave field inhomogeneities, demonstrating the robustness of the pulse.}
 \label{fig:OptimizedPulseFull}
\end{figure*}

\begin{figure*}[p]
 \centering
 \includegraphics[width=12cm]{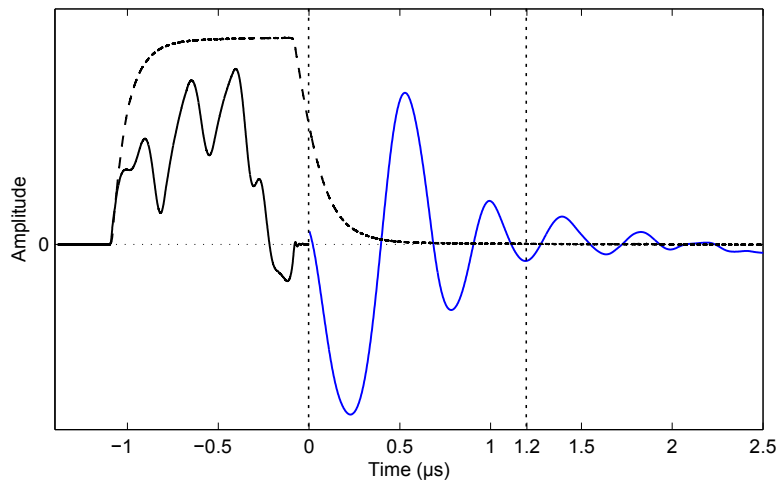}
 \caption{\textbf{Digitized pulse and free-induction decay (FID) of irradiated quartz.} The transients optimized OCT pulse shown in the previous figure was applied to an inhomogeneously broadened solid-state sample of irradiated fused-quartz in a rectangular cavity with Q = 8,486 at a resonance frequency of 9.5236 GHz. The black solid line for $t<0$ shows the pulse profile digitized through a pick-up coil inserted in the cavity. The digitized profile closely matches the calculated profile with the only ringdown being a small oscillation which decays after roughly 75 ns. The resulting quartz FID is shown as a blue solid line for $t>0$ and was acquired after a 75 ns spectrometer deadtime. The static field was moved roughly 2 MHz off-resonance (still within the high-fidelity operation regime of the pulse, as shown in Figure \ref{fig:OptimizedPulseResponse}) to emphasize the shape of the FID. For comparison, a digitized square pulse of length 1 $\mu s$ is shown as the black dashed line. In a separate measurement, the necessary spectrometer deadtime in the absence of ringdown suppression was determined to be 1.2 $\mu s$ (shown as a dotted vertical line), which would prevent the detection of a significant portion of the FID. The separate plots have been scaled for visual clarity.}
 \label{fig:QuartzFID}
\end{figure*}

The pulse was defined piecewise constant over 100 steps of 10 ns each, giving a total length of 1 $\mu$s. The nominal Rabi frequency, $\omega_{1,\mathrm{nom}}/2\pi$ = 5.26 MHz, was determined by Fourier transforming the result of a standard spin-echo Rabi oscillation experiment at a microwave power of 4 Watts and identifying the dominant frequency. The pulse was optimized over a microwave inhomogeneity of $\omega_1 = (0.95, 1, 1.05)$ and a static field inhomogeneity of $\Delta \omega/2\pi$ = $\{$-2 MHz, 2 MHz$\}$ in steps of 250 kHz. The pulse optimization took roughly ten minutes on a standard laptop computer and resulted in an average fidelity over the distribution of $\Phi=0.9905$. The resulting pulse profile and response over an extended distribution is shown in Figure \ref{fig:OptimizedPulseFull}. The free-induction decay (FID) of irradiated fused-quartz resulting from application of the pulse in the Varian cavity is shown in Figure \ref{fig:QuartzFID}, along with the digitized pulse profile measured through the pick-up coil. A spectrometer deadtime of 75 ns was required to allow the small oscillations shown at the end of the digitized pulse to decay. This deadtime was included in the pulse optimization as an additional control period of zero amplitude. These measurements confirm the expected profile of the bandwidth-limited pulse and the ability to observe the FID at high Q. \footnote{Upon initially implementing the optimized pulse we could not observe an FID due to spurious ringdown from a source which did not appear in the waveform digitized through the pick-up coil. The source of this ringdown was determined to be on-resonant leakage of the carrier signal through a double-balanced mixer used to mix in the ideal pulse waveform from an Arbitrary Waveform Generator (AWG). Leakage of this type has been noted previously \cite{Tseitlin2011} and was compensated using a similar method. Before the mixer we split off the carrier signal, phase-shifted it 180 degrees, then recombined it with the amplifier output through a directional coupler, appropriately adjusting the amplitude to cancel the amplified leakage signal. This method of active feedback suppression is also similar to the technique used by Broekaert and Jeener to compensate for radiation damping effects \cite{Broekaert:1995a}.} 

\section{Controllability With Limited Bandwidth}

For accurate spectroscopy and high-fidelity quantum information processing, control sequences must drive transitions and excite coherence over a range of frequencies given by the coupling structure and any uncertainties of the system Hamiltonian. A common solution, based on linear response theory \cite{Kubo1954}, is to require the Fourier spectrum of the control pulse to contain significant contributions from all frequencies present in the quantum system \cite{Morris:1978a,Hoult:1979a}. The principle, and limitations, of this approximate solution may be succinctly demonstrated using a description of the pulse trajectories in $k$-space \cite{Sodickson:1998a}. In this description, each frequency $\omega$ present in the drift Hamiltonian of the quantum system has an associated wavenumber, $k=\omega t$, whose value is modulated by the control Hamiltonian during the length of the applied pulse.

The accuracy of predicting spin response through Fourier analysis depends primarily on the validity of two approximations. The first is that any rotations that occur during a period where the pulse waveform is constant are small, such that the full spin response given by the exponential of the Hamiltonian generating the motion may be truncated to first order. The second is that the amplitude of the control Hamiltonian during any pulse period be significantly greater than the amplitude of the drift Hamiltonian, such that the axis of rotation may be taken as being completely determined by the control Hamiltonian. When these two criteria are satisfied, excitations of the quantum system as a function of frequency may be written simply as the Fourier transform of the time-dependent control amplitudes, $x(t)$:
\begin{equation}
S(\omega) = -i\int_0^{t_p}x(t)e^{-ik} \, \mathrm{d}t,
\end{equation}
where $t_p$ is the pulse length and $S(\omega)$ is the observable spin response generated by the controls, which are taken for simplicity to be amplitude modulated only. If we now write the controls as a sum of Fourier components,
\begin{equation}
x(t) = \sum_{n=-\infty}^{\infty} x_n e^{i \omega^p_n t},
\end{equation}
we see that the only contributions to the observable signal are from Fourier components, $n$, where $x_n \neq 0$, such that the corresponding vector in $k$-space may be effectively refocused during the pulse. Within the linear response approximation, then, pulse design is frustrated by the narrow-band filtering property of a high-Q resonator. However, the ability to numerically optimize pulses allows us to move into a regime where linear response is no longer valid.

For general rotations there is significant mixing of the various $k$-vectors that leads to a complex spin response. Also, a general axis of rotation is not given entirely by the control Hamiltonian, but by the vector sum of the control and drift Hamiltonians. These effects are often small, but may be used to generate spin response deemed inaccessible by linear response. By solving the equations of motion exactly under an accurate system model, without making approximations, we can find bandwidth-limited pulses that retain the same degree of controllability as for infinite bandwidth pulses, albeit with a reduction in efficiency due to the inability to directly address all transitions in the system. 

We first consider an ensemble of uncoupled spins with Larmor frequency, $\omega_0$, off-resonance an amount $\Delta \omega = \omega_0 - \omega_t$ from a control field applied at $\omega_t$. The Hamiltonian in a frame rotating at $\omega_t$ is given by eq. (\ref{eq:SingleSpinHam}), where the term proportional to $\sigma_z$ is the drift Hamiltonian, $H_d$, and the term proportional to $\sigma_x$ is the control Hamiltonian, $H_c$. Generating a computationally universal set of quantum operations requires the ability to generate all elements of the Lie algebra spanning the Hilbert space \cite{DAlessandroBook}. For uncoupled spins, the Lie algebra is SU(2), with basis operators $\left\{I,\sigma_x,\sigma_y,\sigma_z\right\}$. The control algebra generated by a given set of Hamiltonians may be computed by taking successive Lie brackets to all orders \cite{Ramakrishna1995,Schirmer2001}. For $H_d$ and $H_c$ considered here $\left[H_d,H_c\right]\propto \sigma_y$, such that the generated control algebra is identical to SU(2), indicating universal control of the system by appropriate application of the given Hamiltonians. 

The efficiency of control is difficult to quantify in general, but may be posed as the non-commutativity of effective Hamiltonians that may be generated during the pulse between incremental time periods. For SU(2) we may take a simple geometric view, posing the problem of efficiency as the maximum angle of rotation that may be generated over an incremental time period or, equivalently, the maximum angle between effective Hamiltonian vectors that may be generated from one instant in time to the next. Assuming an exponential model of the pulse transients for simplicity, and taking $A(t_0)=0$ and $\omega_1=1$, we examine the non-commutativity of
\begin{equation}
H_{\mathrm{eff}}(t_0) = \frac{1}{2} \Delta \omega \sigma_z,
\end{equation}
and
\begin{equation}
H_{\mathrm{eff}}(t_0+\delta t) = \frac{1}{2} \Delta \omega \sigma_z + \frac{1}{2} A(t_0+\delta t)(1-e^{-\frac{\omega_0}{Q}\delta t}) \sigma_x.
\end{equation}
The controllability of the system has not changed, provided $Q$ is finite, as these Hamiltonians generate the same Lie algebra as for infinite bandwidth. The angle between the two Hamiltonian vectors, $\theta$, is given by
\begin{equation}
\theta = \tan^{-1}\left[\frac{A(t_0+\delta t)}{\Delta\omega}(1-e^{-\frac{\omega_0}{Q}\delta t})\right].
\end{equation}
The size of the generated incremental rotation depends on the ratio of the target control amplitude to the resonance offset and the value of $Q$, with increasing efficiency for larger control amplitudes and smaller $Q$. Given that control amplitude is proportional to $\sqrt{Q}$, there is an inherent trade-off between sensitivity and the efficiency of generating rotations that must be considered. However, given sufficient time and a sufficiently accurate system model, pulses may be optimized that address transitions that are arbitrarily far off-resonance without containing Fourier components anywhere near the transition frequency. In the following section we provide examples, in the context of anisotropic hyperfine coupled electron-nuclear spin systems, of pulses that significantly exceed the limitations imposed by linear response theory.

\section{Application of Bandwidth-Limited Control to Electron-Nuclear Spin Systems}
We consider a one electron, one nuclear (1e-1n) spin system in a strong external magnetic field $\vec{B_0} = B_0 \hat{z}$ with an anisotropic hyperfine interaction coupling the two spins. The drift Hamiltonian is given by
\begin{equation}
H_d = \frac{1}{2}\omega_z^e \sigma_z^e + \frac{1}{2}\omega_z^n \sigma_z^n + \frac{1}{4} \omega_{zz} \sigma_z^e\sigma_z^n + \frac{1}{4} \omega_{zx}\sigma_z^e\sigma_x^n,
\end{equation}
where $\omega_z^e$ and $\omega_z^n$ are the strengths of the electron and nuclear Zeeman interactions, respectively, with the static field, and $\omega_{zz}$ and $\omega_{zx}$ are isotropic and anisotropic components of the hyperfine interaction. Examples of such systems include solid-state organic molecules with a localized free radical \cite{McCalley:1993a,Heidebrecht:2006a} and nitrogen-vacancy defect centers in diamond \cite{Jelezko:2006a}. 

The electron Zeeman interaction is dominant and defines the principle coordinate system of spin quantization. In terms of this coordinate system, the eigenstates of the nuclear spins are given by the vector sum of the nuclear Zeeman interaction and the spin dependent local field generated by the anisotropic hyperfine coupling to the electron:
\begin{equation}
\begin{split}
& \ket{1} = \ket{\uparrow \alpha_0} = \sin\theta_{\uparrow}\ket{\uparrow\uparrow} + \cos\theta_{\uparrow}\ket{\uparrow\downarrow}, \\
& \ket{2} = \ket{\uparrow \alpha_1} = \cos\theta_{\uparrow}\ket{\uparrow\uparrow} - \sin\theta_{\uparrow}\ket{\uparrow\downarrow}, \\ 
& \ket{3} = \ket{\downarrow \beta_1} = \cos\theta_{\downarrow}\ket{\downarrow\uparrow} - \sin\theta_{\downarrow}\ket{\downarrow\downarrow}, \\
& \ket{4} = \ket{\downarrow \beta_0} = \sin\theta_{\downarrow}\ket{\downarrow\uparrow} + \cos\theta_{\downarrow}\ket{\downarrow\downarrow},
\end{split}
\end{equation}
where $\theta_{\uparrow}$ and $\theta_{\downarrow}$ determine the non-commutativity of the resulting eigenstates and are given by:
\begin{equation}
\begin{split}
& \theta_{\uparrow} = \tan^{-1}\left(\frac{-\omega_{zx}}{\omega_{zz}-\omega_z^n}\right), \\
& \theta_{\downarrow} = \tan^{-1}\left(\frac{-\omega_{zx}}{\omega_{zz}+\omega_z^n}\right).
\end{split}
\end{equation}

Due to the non-commutativity of the local fields seen by the nuclear spin when the electron spin is in the spin-up versus spin-down state, universal control over the entire spin system (generalizing to 1e-Nn systems) may be achieved by implementing only the electron $\sigma_x^e$ generator and allowing free evolution under $H_d$ \cite{Hodges2008,Khaneja2007}. This may also be seen by noting that the transition probability between any pair of non-degenerate states associated with the $\sigma_x^e$ operation is non-zero \cite{Schweiger1995}. We may thus use the nuclear spins as a quantum processor, with the electron spin acting as an actuator element to allow for fast quantum operations on the processor and providing a means for efficiently transferring information in a node-based quantum information processor design \cite{Borneman:2012a,Zhang2011,Hodges2008}.

To demonstrate that universal control via electron-only modulation may still be achieved when using a high-Q resonator we numerically optimized a pulse which implements an electron spin flip,
\begin{equation}
U_d = e^{-i \frac{\pi}{2} \sigma_x^e \otimes I^n},
\end{equation} 
for a 1e-1n sample in a resonator with bandwidth smaller than the size of the hyperfine interaction. The ability to achieve such an operation implies full controllability of the spin system with electron-only modulation. The control Hamiltonian was taken to be a time-dependent amplitude modulation of the electron spin only,
\begin{equation}
H_c(t) = \frac{1}{2} A(t) \sigma_x^e \otimes I^n.
\end{equation}

\begin{figure*}[p]
 \centering
 \subfloat[Pulse Profile]{\label{fig:ElecNucPulseProfile}\includegraphics[width=12cm]{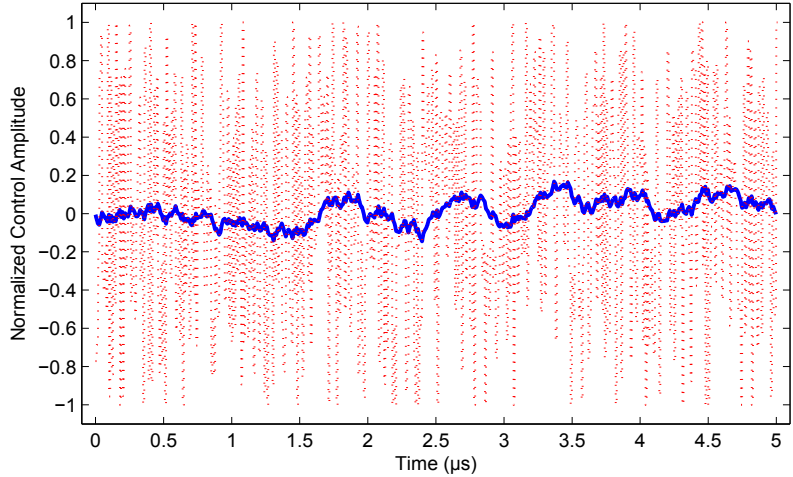}}
 \\
 \subfloat[Pulse Fourier Spectrum]{\label{fig:ElecNucPulseFT}\includegraphics[width=12cm]{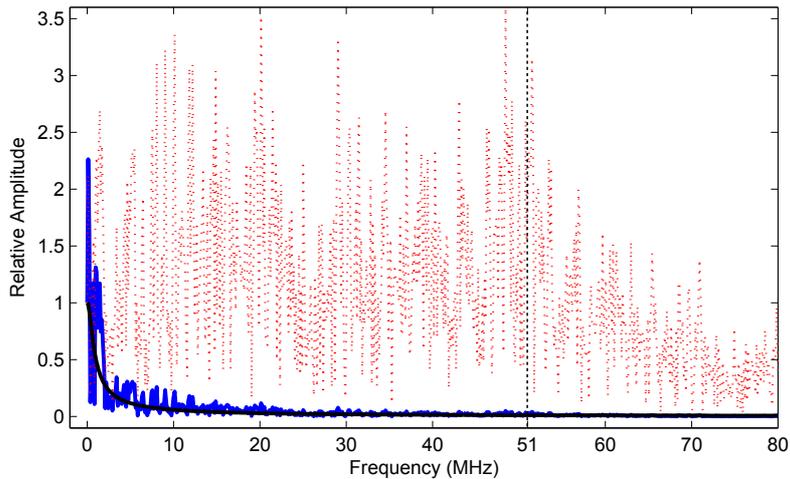}}
 \caption{\textbf{Optimized 1e-1n pulse at high-Q.} (a) The undistorted (dashed line) and distorted (solid line) control amplitudes of a pulse designed to perform a $\pi$ rotation of the electron spin in a resonator with Q = 10,000. Access to such an operation guarantees universal control of the nuclear spin via electron-only modulation \cite{Hodges2008}. The pulse consists of 500 time steps of $\Delta$t= 10 ns each for a total length of 5 $\mu$s, with $\omega_{1,\mathrm{nom}}/2\pi$ = 100 MHz. The simulated average gate fidelity is $\Phi$ = 0.9901. (b) The single-sided amplitude spectrum of the undistorted pulse (dotted line) and the distorted pulse (solid line) filtered by the resonator admittance function (bold solid line). The transitions necessary to achieve the desired operation were separated by an amount ($\left|\omega_{23}-\omega_{14}\right|$ = 51 MHz) much greater than the bandwidth of the resonator ($\approx$ 1.2 MHz), demonstrating that linear response may be greatly exceeded by pulses of high complexity optimized under a sufficiently accurate system model.}
 \label{fig:HighQPulseFull}
\end{figure*}

The pulse was optimized using the algorithm outlined in \ref{Sec:DistortedControls} using a model of the resonator given in \ref{Sec:ResonatorModel} with a Q of 10,000 (BW $\approx$ 1.2 MHz). The drift Hamiltonian parameters were taken from \cite{Hodges2008}: $\omega_z^e/2\pi = 11.885$ GHz, $\omega_z^n/2\pi = 18.1$ MHz, $\omega_{zz}/2\pi = -42.7$ MHz, and $\omega_{zx}/2\pi = 14.2$ MHz. The carrier frequency of the pulses was set resonant with the 1-4 transition, $\omega_0/2\pi = 11.909$ GHz. The nominal Rabi frequency was taken to be $\omega_{\mathrm{nom}}/2\pi = 100$ MHz. The resulting pulse is shown in Figure \ref{fig:ElecNucPulseProfile}. A pulse time of 5 $\mu$s was chosen for convenience to provide sufficient time to easily achieve the desired operation. We expect solutions for shorter pulse times to exist, but have not systematically addressed the minimum time needed to achieve the operation for a given set of parameters. The final simulated average gate fidelity of the pulse was 0.9901 even when, as shown in Figure \ref{fig:ElecNucPulseFT}, all significant frequency modulation is much less than the hyperfine splitting. 

\section{Conclusions}
Integrating the resonator impulse response function and ringdown suppression into an optimal control theory pulse design algorithm allows the use of high-Q resonators in inductive measurements, enabling higher signal-to-noise ratio and sensitivity without sacrificing the ability to perform a universal set of quantum operations with high fidelity. These techniques also permit the direct observation of the free-induction decay of inductively detected samples at high Q, which was experimentally demonstrated on a sample of irradiated fused-quartz in a high-Q rectangular cavity. We focused on a particular application of the described methods in pulsed electron spin resonance where universal control of the quantum state of a nuclear spin may be achieved by microwave-only modulation of an anisotropic hyperfine coupling to an electron spin. We demonstrated that the limits imposed by linear response theory may be vastly exceeded when using numerically optimized pulses, allowing universal control of the joint electron-nuclear spin system even when the hyperfine coupling strength is significantly greater than the resonator bandwidth. These solutions rely on the non-commutativity of the effective Hamiltonians generated throughout the pulse and the ability to accurately engineer the exact response of a quantum system to complex modulations.

We expect the methods presented in this work to find application in a broad range of fields. In addition to spectroscopic and imaging applications in chemistry, biology, and materials science for samples with a small number of spins, we expect the application of high-Q control techniques to molecular thin film samples to be particularly compelling due to their relevance to building a large-scale multinode quantum information processor. The described techniques may also find application in hybrid quantum systems that aim to use spin ensembles as memory elements for microwave photons \cite{Wesenberg:2009a,Kubo:2012a}. 

We note that the optimized pulses work well only for the specific set of system parameters defined at the time of optimization and the performance is sensitive to any change or mis-setting of the system parameters when running an experiment. The pulses may be made robust to variations in system parameters, but with a corresponding increase in pulse time and complexity \cite{Borneman:2010a,Boulant2004,Pravia2003}. The efficiency of control when both robustness and high-fidelity are required of pulses to be used with a high-Q resonator requires further investigation.

\section{Appendix: Derivation of Distorted Gradients}
We begin with a set of undistorted controls, $\{u^j\}$, where $j = 1,...,N$ and resample to obtain a set of undistorted controls $\{\widetilde{u}^m\}$, where $m = 1,...,M$ and $M=n_s N + n_r$ for $n_s$ samples per $N$ control periods and $n_r$ values of zero appended to the waveform to account for pulse ringdown. For clarity, we assume only one control Hamiltonian ($k=1$). The distorted controls, $\{\widetilde{v}^m\}$, are then given by the discrete convolution of the undistorted controls with the resonator impulse response function:
\begin{equation}
\widetilde{v}^m = \sum_{l\leq m}{h^{m-l+1}\widetilde{u}^l}.
\end{equation}
The average gate fidelity may then be written as
\begin{equation}
\Phi = \left<U_d|\widetilde{U}_M...\widetilde{U}_1\right>\left<\widetilde{U}_1...\widetilde{U}_M|U_d\right>,
\end{equation}
where the $m^{\mathrm{th}}$ propagator is calculated from the distorted controls as
\begin{equation}
\begin{split}
\widetilde{U}_m = &e^{-i\frac{\Delta t}{n_s}(H_d+\tilde{v}^m H_1)}\\
& = e^{-i \frac{\Delta t}{N_s} (H_d + \sum_{l\leq m}h^{m-l+1}\tilde{u}^{m}H_1)}.
\end{split}
\label{eq:ResampledPropagator}
\end{equation}
We are interested in calculating the gradient of the fidelity with respect to a perturbation of the undistorted controls $\{u^j\}$, given by the usual product and chain rules of derivation as
\begin{equation}
\begin{split}
\frac{\delta\Phi}{\delta u^j} = & \sum_{m=1}^M\left<U_d|\widetilde{U}_M...\frac{\delta \widetilde{U}_m}{\delta u^j}...\widetilde{U}_1\right>\left<\widetilde{U}_1...\widetilde{U}_M|U_d\right> + \\ & \left<U_d|\widetilde{U}_M...\widetilde{U}_1\right>\left<\widetilde{U}_1...\frac{\delta \widetilde{U}_m}{\delta u^j}...\widetilde{U}_M|U_d\right>.
\end{split}
\end{equation}
By defining a forward propagator, $\widetilde{X}_m = \widetilde{U}_m...\widetilde{U}_1$, a backward propagator, $\widetilde{P}_m = \widetilde{U}^{\dag}_{m+1}...\widetilde{U}^{\dag}_M U_d$, and noting that only propagators with $m > (j-1)n_s$ will be modified by a change in the $j^{\mathrm{th}}$ control period, we may simplify this expression to
\begin{equation}
\begin{split}
\frac{\delta\Phi}{\delta u^j} = & \sum_{m > (j-1)n_s}^M \left<P_m|X_m\right>\left<\frac{\delta \widetilde{U}_m}{\delta u^j}X_{m-1}|P_m\right> + \\ & \left<P_m|\frac{\delta \widetilde{U}_m}{\delta u^j}X_{m-1}\right>\left<X_m|P_m\right>.
\end{split}
\label{eq:RawGradient}
\end{equation}
Due to resampling the control period, we define a top-hat function, $\Xi(j,n_s)$, to account for the perturbation being constant during the entire $j^{\mathrm{th}}$ control period:
\begin{equation}
\Xi(j,n_s) = 1 \, \mathrm{for} \, (j-1)n_s < m \leq j n_s + 1,
\end{equation} 
\begin{equation}
\Xi(j,n_s) = 0 \, \mathrm{otherwise}. 
\end{equation} 
The derivative $\delta \widetilde{U}_m/\delta u^j$ may then be calculated by definition as
\begin{equation}
\frac{\delta \widetilde{U}_m}{\delta u^j} = \lim_{\delta u^j \rightarrow 0} \frac{\widetilde{U}_m(\widetilde{v}^m+ \Xi(j,n_s) \delta u^j)-\widetilde{U}_m(\widetilde{v}^m)}{\delta u^j}.
\label{eq:DerivativeDefinition}
\end{equation} 
We apply a small perturbation to the $j^{\mathrm{th}}$ control, $u^j \rightarrow u^j+\delta u^j$ and determine the resulting $m^{\mathrm{th}}$ propagator:
\begin{equation}
\begin{split}
& \widetilde{U}_m(\widetilde{v}^m+\Xi(j,n_s) \delta u^j) = \\ &
 e^{-i\frac{\Delta t}{n_s} \left[H_d + (\widetilde{v}^m + \sum_{l \leq m} h^{m-l+1} \Xi(j,n_s) \delta u^j) H_1\right]}.
\end{split}
\end{equation}
Assuming $\Delta t$ to be small, we may approximate $H_d$ and $H_1$ as commuting, allowing us to keep only the first order term of a BCH expansion of the perturbed propagator
\begin{equation}
\begin{split}
& \widetilde{U}_m(\widetilde{v}^m+\Xi(j,n_s)u^j) = \\ &
\widetilde{U}_m(\widetilde{v}^m) e^{-i\frac{\Delta t}{n_s} \sum_{l \leq m} h^{m-l+1} \Xi(j,n_s) \delta u^j H_1}.
\end{split}
\end{equation}
If we now assume the perturbation $\delta u^j$ to also be small, we may make a first order series approximation of the exponential, to give
\begin{equation}
\begin{split}
& \widetilde{U}_m(\widetilde{v}^m+\Xi(j,n_s)u^j) = \\ &
\widetilde{U}_m(\widetilde{v}^m)(I -i \frac{\Delta t}{n_s} H_1 \widetilde{U}_m \, \sum_{l \leq m} h^{m-l+1} \Xi(j,n_s) \delta u^j + \mathcal{O}(\Delta t^2).
\end{split}
\end{equation}
We may now plug this expression back into (\ref{eq:DerivativeDefinition}) to obtain
\begin{equation}
\frac{\delta \widetilde{U}_m}{\delta u^j} = -i \frac{\Delta t}{n_s} H_1 \widetilde{U}_m \, \sum_{l \leq m} h^{m-l+1} \Xi(j,n_s),
\end{equation}
which, when plugged into (\ref{eq:RawGradient}) gives
\begin{equation}
\frac{\delta\Phi}{\delta u^j} = \sum_{m > (j-1)n_s}^M \xi^m(j) \frac{\delta\Phi}{\delta \widetilde{v}^m},
\end{equation}
where $\xi^m(j)$ is the convolution of the top-hat function with the resonator impulse response function:
\begin{equation}
\xi^m(j) = \sum_{l \leq m} h^{m-l+1} \Xi(j,n_s).
\end{equation}
This function may be interpreted as a weighting of the change of the fidelity due to a small change $\delta \widetilde{v}^m$ of the pulse parameter at the $m^{\mathrm{th}}$ evolution time step induced by the perturbation $\delta u^j$:
\begin{equation}
\frac{\delta\Phi}{\delta \widetilde{v}^m} = -2\mathrm{Re}\left[\left<P_m|i\frac{\Delta t}{n_s} H_1 X_m\right>\left<X_m|P_m\right>\right].
\end{equation} 
Note that when the controls are undistorted, the gradient weighting function is equivalent to the top-hat function, giving
\begin{equation}
\frac{\delta\Phi}{\delta u^j} = \sum_{m \geq (j-1)n_s +1}^{j n_s} \frac{\delta\Phi}{\delta \widetilde{u}^m},
\end{equation}
which for short time steps and small perturbations, reduces to the undistorted gradient derived in \cite{GRAPE2005}:
\begin{equation}
\frac{\delta\Phi}{\delta u^j} = -2\mathrm{Re}\left[\left<P_j|i\Delta t H_1 X_j\right>\left<X_j|P_j\right>\right].
\end{equation}

In the special case where two control Hamiltonians, $H_1$ and $H_2$, are in quadrature with one another - for example, amplitude and phase modulation of a single control field in magnetic resonance - the gradients must be modified to reflect the interdependence of the Hamiltonians. The resonator impulse response function, $h$, and the control signals, $u$ and $v$, must be considered as complex functions, with the real part associated with $H_1$ and the imaginary part associated with $H_2$. The gradients for the real and imaginary parts of the controls are then:
\begin{equation}
\frac{\delta\Phi}{\delta \mathrm{Re}[u^j]} = \sum_{m > (j-1)n_s}^M \mathrm{Re}[\xi^m(j)] \frac{\delta\Phi}{\delta \mathrm{Re}[\widetilde{v}^m]} + \mathrm{Im}[\xi^m(j)] \frac{\delta\Phi}{\delta \mathrm{Im}[\widetilde{v}^m]},
\end{equation}
\begin{equation}
\frac{\delta\Phi}{\delta \mathrm{Im}[u^j]} = \sum_{m > (j-1)n_s}^M \mathrm{Re}[\xi^m(j)] \frac{\delta\Phi}{\delta \mathrm{Im}[\widetilde{v}^m]} - \mathrm{Im}[\xi^m(j)] \frac{\delta\Phi}{\delta \mathrm{Re}[\widetilde{v}^m]},
\end{equation}
where
\begin{equation}
\frac{\delta\Phi}{\delta \mathrm{Re}[\widetilde{v}^m]} = -2\mathrm{Re}\left[\left<P_m|i\frac{\Delta t}{n_s} H_1 X_m\right>\left<X_m|P_m\right>\right],
\end{equation} 
\begin{equation}
\frac{\delta\Phi}{\delta \mathrm{Im}[\widetilde{v}^m]} = -2\mathrm{Re}\left[\left<P_m|i\frac{\Delta t}{n_s} H_2 X_m\right>\left<X_m|P_m\right>\right].
\end{equation}

\section{Acknowledgements} 
This work was supported by the Canadian Excellence Research Chairs (CERC) Program and the Canadian Institute for Advanced Research (CIFAR). We also thank Jeremy Chamilliard, Chandrasekhar Ramanathan, Hamid Mohebbi, Clarice Aiello, Jonathan Hodges, and Holger Haas for helpful discussions.

\bibliographystyle{elsarticle-num}
\bibliography{References_Borneman}

\end{document}